\documentclass[a4paper,fleqn,usenatbib]{mnras}

\usepackage[T1]{fontenc}
\usepackage{ae,aecompl}

\usepackage{graphicx}	
\usepackage{amsmath}	
\usepackage{amssymb}	

\usepackage{textcomp}
\usepackage{cleveref}
\newcommand{\mr}{\mathrm}
\newcommand{\sm}{\hbox{$M_\odot$}}
\newcommand{\sr}{\hbox{$R_\odot$}}
\newcommand{\Bpara}{\hbox{$B_\parallel$}}

\newcommand{\DMunit}{\hbox{${\rm pc/cm}^3$}}

\usepackage{xcolor} 

\title[Magnetic field in binary interface]{
  Constraining small scale magnetic fields through plasma lensing:
  Application to the Black widow eclipsing pulsar binary
}

\author[D. Li et al.]{
Dongzi Li$^{1,2,3}$\thanks{E-mail: dzli@cita.utoronto.ca},
Fang Xi Lin$^{1,2}$,
Robert Main$^{1,4,3}$,
Ue-Li Pen$^{1,5,3,6}$,
\newauthor
Marten H. van Kerkwijk$^{4}$, and
I-Sheng Yang$^{1,6}$
\\
$^{1}$Canadian Institute for Theoretical Astrophysics, University of Toronto, 60 St. George Street, Toronto, ON M5S 3H8, Canada\\
$^{2}$Department of Physics, University of Toronto, 60 St. George Street, Toronto, ON M5S 1A7, Canada\\
$^{3}$Dunlap Institute for Astronomy and Astrophysics, University of Toronto, 50 St. George Street, Toronto, ON M5S 3H4, Canada\\
$^{4}$Department of Astronomy and Astrophysics, University of Toronto, 50 St. George Street, Toronto, ON M5S 3H4, Canada\\
$^{5}$Canadian Institute for Advanced Research, 180 Dundas St West, Toronto, ON M5G 1Z8, Canada\\
$^{6}$Perimeter Institute for Theoretical Physics, 31 Caroline Street North, Waterloo, ON N2L 2Y5, Canada\\
}

\date{Accepted XXX. Received YYY; in original form ZZZ}

\pubyear{2018}

\begin{document}
\label{firstpage}
\pagerange{\pageref{firstpage}--\pageref{lastpage}}
\maketitle

\begin{abstract}
In regions with strongly varying electron density, radio emission can
be magnified significantly by plasma lensing.
In the presence of magnetic fields, magnification in time and frequency will be different for two circular polarizations.
We show how these effects can be used to measure or constrain the magnetic field  parallel to the line of sight,
$B_\parallel$, as well as its spatial structure,
$\sigma_{B_\parallel}$, in the lensing region. In addition, we discuss how generalized
Faraday rotation can constrain the strength of the perpendicular field,
$B_\perp$.  We attempt to make such measurements for the Black Widow
pulsar, PSR~B1957+20, in which plasma lensing was recently discovered.
For this system, pressure equilibrium suggests
$B\gtrsim 20\,$G at the interface between the pulsar and companion
winds, where the radio eclipse starts and ends, and where most lensing
occurs. We find no evidence for large-scale magnetic fields, with,
 on average, $B_\parallel=0.02\pm0.09\,$G over the egress lensing region.
From individual lensing events, we strongly constrain small scale
magnetic structure to $\sigma_B<10\,$mG, thus excluding
scenarios with a strong but rapidly varying field.  Finally, from the
lack of  reduction of average circular polarization in
the same region, we rule out a strong, quasi-transverse field.
We cannot identify any plausible scenario in which a large magnetic
field in this system is concealed, leaving
the nature of the interface between the pulsar and companion winds an enigma.
Our method can be applied to other sources showing plasma lensing,
including other eclipsing pulsars and fast radio bursts,
to study the local properties of the magnetic field.
\end{abstract}



\section{Introduction}
The success of gravitational lensing shows that
a remarkable amount of information about the material distribution
along the line of sight can be gleaned from the bending and delaying of light
(eg. \citealt{92Schneider}).
At radio frequencies, lensing by plasma can similarly illuminate structures in the interstellar medium (see, e.g., \citealt{16Tuntsov}).
Unlike gravitational lenses,
plasma lenses are highly chromatic, as the refractive index of a
plasma depends on the frequency of light.
Moreover, in the presence of magnetic fields, the refractive indices of the left and right circular polarization states differs,
a phenomenon known as birefringence.
Birefringence is sensitive to spatial structure of the magnetic fields, and completely predictable in the case of a uniform magnetic field.  It
thus serves as a probe for magnetic properties at
the location of lenses.

For conditions typical of the interstellar medium, with magnetic field strengths of order tens of $\mu$G, the effects are relatively small, but stronger fields are encountered in supernova remnants and in binary systems in which the companion is loosing mass.
Of particular interest here are the so-called ``Black Widow'' and ``Redback'' pulsar binaries,
in which the pulsar signal is seen to be dispersed or removed by plasma associated with the companion.
This plasma also can lens the pulsar, as recently found for the original Black Widow binary, PSR B1957+20, where in regions near the radio eclipse the pulses not only show variations in dispersive delay of order 1\,$\mu$s on a timescale of 2\,s, but also in amplitude, with magnifications reaching nearly two orders of magnitude \citep{18Main}.
The variable magnifications are best explained through plasma lensing, and here we show how the flux modulation
versus time, frequency and polarization from the brightest events
helps constrain the magnetic properties of these plasma lenses.


The paper is organized as follows.
In Section \ref{sec:motivation}, we first discuss the properties of PSR B1957+20 and its eclipse,
in particular the need of a strong magnetic field to understand the long radio eclipse, and the
tension between this expected strong field with previous observational upper limits to the field strength.
Next, in Section \ref{sec:effects}, we illustrate the observable effects of magnetic fields,
introducing quantitive methods to use birefringence of plasma lenses,
and reviewing methods beyond Faraday rotation. We apply these methods in
Section \ref{sec:observations}, separating what we can learn from normal, lensed, and giant pulses.
We infer substantially more stringent upper limits on the magnetic field strength, and
discuss their impact on our understanding of the system in Section \ref{sec:alternatives}.
Finally, in Section \ref{sec:conclusion}, we present the ramifications of our results.

\section{Motivation for measuring the magnetic field}
\label{sec:motivation}
PSR B1957+20 was the first millisecond pulsar binary found in which the pulsar signal
is eclipsed near superior conjunction of its companion \citep{88Fruchter}.
Thirty years after its discovery, many puzzles remain.  Most relevant for our purposes
are two associated with the eclipse.
First, the excess electrons observed near eclipses implies material extending to a distance of
$r_\mr{if}\simeq0.8\,\sr$
from the companion in the direction perpendicular to the pulsar wind, which is well outside the companion's Roche lobe;
what supports this material against the strong Poynting flux from the pulsar?
Second, observations at higher frequency show that when the radio flux
is greatly reduced at 318 MHz, the excess electron column density is of order $\sim0.01\,$\DMunit \citep{91Ryba}.
For a typical length scale of the system of $\sim\!0.5\,\sr$, this
indicates an electron density of only $n_e \sim 10^6{\rm\,cm^{-3}}$, much lower
than required for simple absorption mechanisms like free-free absorption;
what then is the mechanism causing the long radio eclipse?

\citet{94Thompson} suggested a simple solution to both puzzles, viz., that the companion had a
magnetic field which was strong enough to lead to a field of 20--40\,G at the interface
between the companion's outflow and the pulsar wind. If so,
the magnetic pressure would suffice to balance the Pointing flux, thus explaining
the long stand-off distance,
and synchrotron-cyclotron absorption would suffice to explain the radio eclipse.
Measurements by \citet{90Fruchter}, however,
constrained the average $\overline{B}_\parallel$ to be
$<\!6\,$G and $<\!1.5\,$G right before and after the radio eclipse, respectively.
Given the discrepancy between these limits and the expected large magnetic fields,
it seems well worthwile to attempt to measure the magnetic field more carefully.

\subsection{The predicted magnetic strength}
If the pulsar wind is balanced by the magnetic pressure of the companion wind at the interface, then
\begin{equation}
	P_\mr{pw}=P_\mr{cw} =\frac{B_\mr{cw}^2}{8\pi}.
\end{equation}
where `pw' indicates pulsar wind, `cw' indicates companion wind.

The pressure of the pulsar wind can be estimated from the spin down energy of the pulsar scaled to the distance of companion:
\begin{equation}
    P_\mr{pw}=\frac{I_\mr{psr}\Omega\dot\Omega}{4\pi a^2c},
\end{equation}
where $I_\mr{psr}$ is the pulsar's moment of inertia,
$\Omega\equiv2\pi/P$ the angular spin frequency,
$P$ the spin period, and
$a$ the orbital separation.
Inserting $a=2.7$\,\sr, $P=1.6$\,ms, and $\dot{P}=1.7\times 10^{-17}$\,ms \citep{94Arzoumanian},
one infers that to withstand the pulsar wind ram pressure requires a magnetic field
strength,
\begin{equation}
	B_{\rm cw} = 19\,G\, I_{45}^{1/2} \gtrsim20\,G,
\end{equation}
where the moment of inertia is scaled to $10^{45}{\rm\,g\,cm^2}$,
and the approximate lower limit includes that, for PSR B1957+20, $I_{45}>1.5$ \citep{15Steiner}
given the lower limit 1.65$\,$\sm\ to its mass \citep{11Marten},
and that the above estimate assumes a spherically symmetric pulsar wind,
while in reality it is probably focused towards the equatorial plane
(in which the companion likely resides).

Given the predicted value of $B_\mr{cw}\gtrsim20\,$G at the interface,
a simple dipole model yields a surface magnetic field of $B_\mr{comp}\simeq B_\mr{cw}(r_\mr{if}/R_\mr{comp})^3\gtrsim650$\,G
(where $R_\mr{comp}\simeq0.25\,\sr$; \citealt{11Marten}).
This is quite plausible, given that the companion is a rapidly rotating brown dwarf,
and brown dwarfs with surface magnetic fields as high as 5\,kG are known \citep{17Berdyugina}.

\subsection{Scenarios for hiding strong fields}
\label{sec:scenarios}
Given the arguments above, there should be a strong, $\sim\!20\,$G field
that provides pressure balance at the interface, yet such a strong field
seems excluded by observational limits.  Could it be that the observations
are not sensitive to the true field strength?

One possibility arises from the measurement technique:
because of the small linear polarization,
the best constraints on $\overline{B}_\parallel$ from \citet{90Fruchter}
were from measurements of the Faraday group delay between left and right circular
polarizations in profiles that were integrated over 10--60\,s.
As argued by \citet{94Thompson}, however, the magnetic field in the pulsar
wind -- i.e., well outside the light cylinder where the field cannot corotate --
should change direction on a length scale of $cP/2 \approx 240$\,km.
As the Poynting flux hits the magnetized companion wind, rapid
reconnection is expected at the interface, and the magnetic field in the
companion wind is thus expected to vary on the same length scale.
If so, for an observation of Faraday delay made through the reconnection layer,
the Faraday delay will be averaged down along the line of sight, and a
further reduction will happen by integrating over multiple pulses.
To avoid this, one would ideally use single-pulse measurements of the
magnetic field, and also obtain constraints on the magnetic variance.
We will show in Section \ref{sec:theory} how birefringence in pulses
magnified by plasma lenses can provide such constraints.

Another simple scenario to square the low $B_\parallel$ observed
with the high expected value is to assume the 20\,G magnetic field is
nearly perpendicular to the line of sight.  As discussed in Section \ref{sec:GFR},
this {\em ad hoc} assumption can, for sufficiently strong magnetic fields,
be tested with generalized Faraday rotation.

Finally, underlying the constraints on the magnetic
field is an assumption that it is companion material that is being traversed,
i.e., that the excess DM represents excess electrons accompanied by ions.
Could it be that the excess material instead consists of
electron-positron pairs, either material captured from the pulsar wind (perhaps
similar to the solar wind captured in Earth's Van Allen belts),
or pairs created by interaction of the pulsar wind with companion material?
In a scenario were the excess DM is due to pairs, no differences in Faraday delay
would be expected, since electrons and positrons have opposite and equal effect.
An advantage of this more extreme scenario is that it might also help understand
the eclipses seen for PSR J1816+4510 \citep{14Stovall}, whose companion is a
proto--white dwarf \citep{13Kaplan} for which, unlike for the low-mass companions
typically found for eclipsing binary pulsars, no mass loss is expected.
As we show below, this scenario can also be tested using generalized
Faraday rotation.

\begin{figure*}
    \begin{minipage}{\linewidth}
        \center
    \includegraphics[width=0.5\textwidth,height=0.64\textwidth]{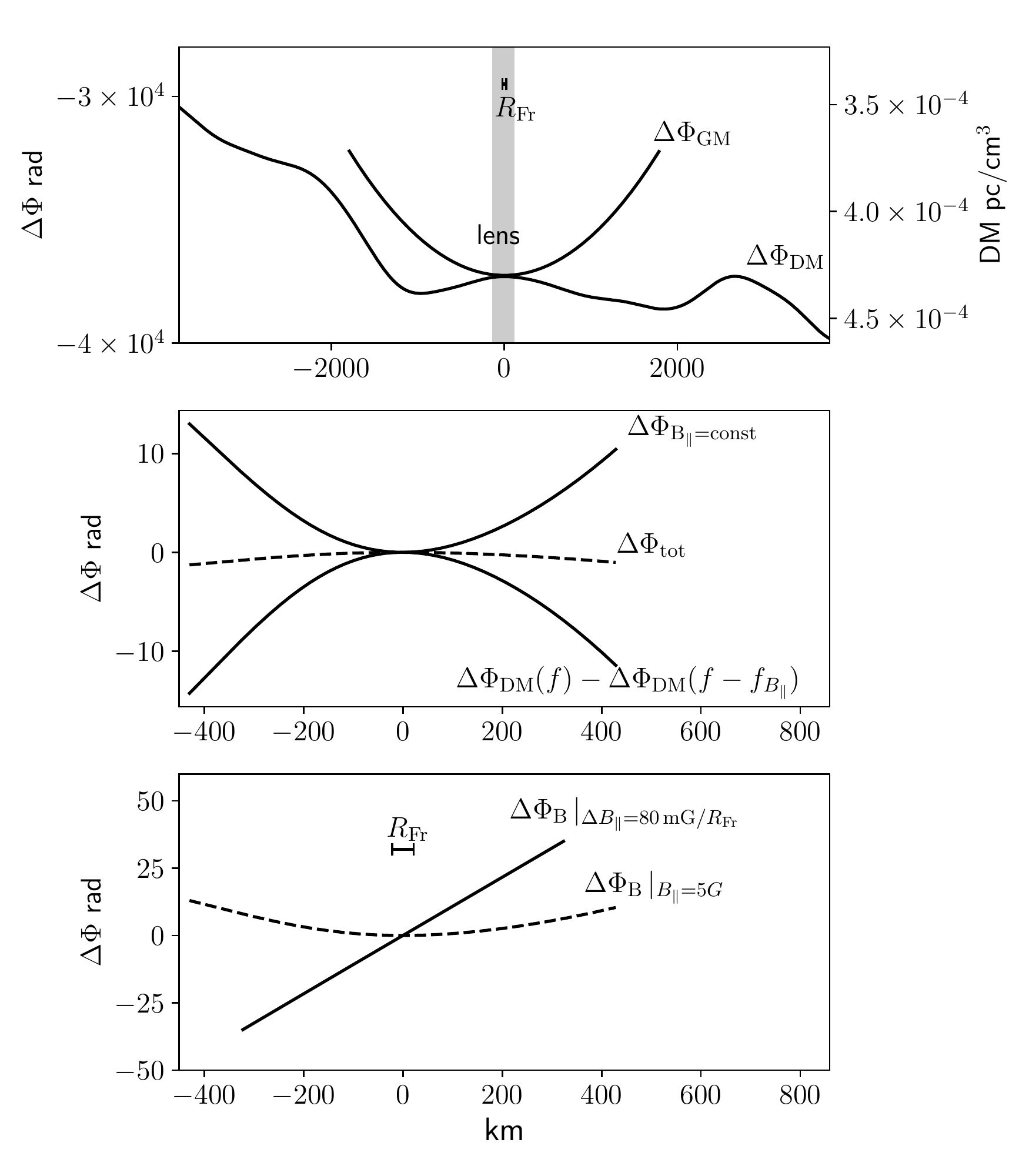}
    \vspace{-0.1cm}
    \hspace{-0.8cm}
    \hfill
    \includegraphics[width=0.5\textwidth,height=0.64\textwidth]{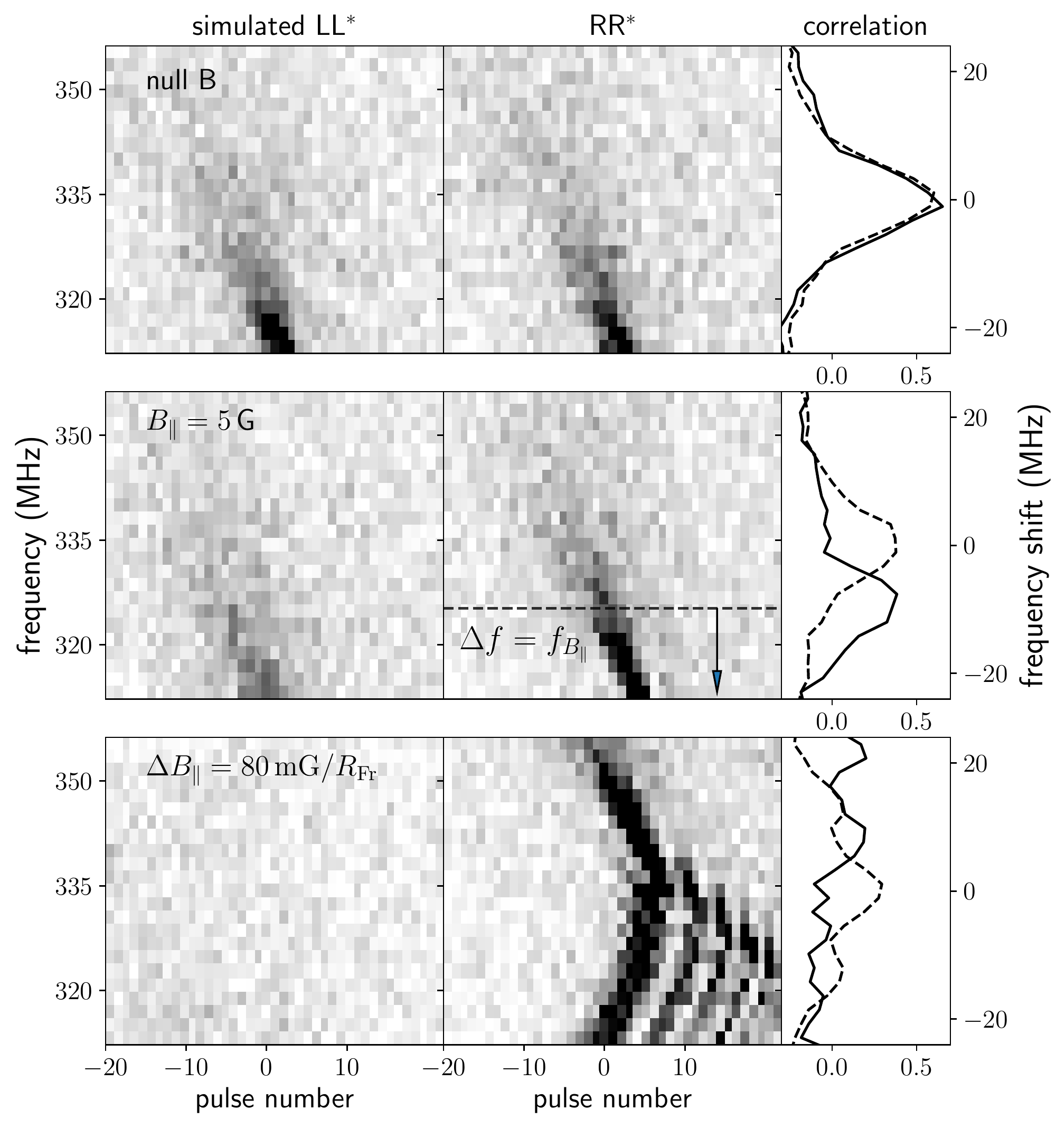}
    \end{minipage}
    \caption{Induced birefringence in plasma lenses due to the influence of magnetic fields
        on the incoming wavefronts for left and right circular polarizations.
        The left panels show wavefront phases for left circular polarization
        (for right polarization, $\Delta\Phi_B$ has the opposite sign),
        while the right panels show sample simulated spectra with cross
    polarization frequency correlation of pulse number
    0 at the side.
        ({\em Top}): Extreme lensing events happen whenever
        the phase delays $\Delta\Phi_\mr{DM}$ caused by excess DM happen to
        cancel out the geometric delays $\Delta\Phi_\mr{GM}$ over regions larger than the Fresnel scale
        (for this example, $R_{\rm Fr}=45\,$km)
        Given the chromaticity of the dispersive delays, the expected spectra are chromatic, and,
        in the absence of magnetic fields, identical for both polarizations within the noise.
        ({\em Middle}): In the presence of a uniform magnetic field in
        the plasma lens, convergence in the two polarization states will happen
        at a different frequency, because the phase gradient induced by a uniform magnetic field
        can be compensated by a shift in frequency.  As a result,
        the spectra of the two polarizations will be offset by the cyclotron frequency $f_{B_\parallel}$.
        ({\em Bottom}): In the presence of magnetic variations in the transverse plane,
        the coherency of the lens can be destroyed.  For a DM of $\sim\!10^{-4}$pc/cm$^3$,
        a magnetic wedge of $\Delta B_\parallel=80\,\mathrm{mG}/R_\mathrm{Fr}$ will completely change the curvature of
        the wavefront, and hence, the two polarizations will not be magnified at the same time.
        (Note that in the simulation for this case, the ringing results from interference between
        multiple images at frequencies right below a caustic. This itself is not a magnetic effect,
        but a modulation that is shifted into the band -- it is present below the parts shown
        also in the other examples.)
    }
\label{fig:simulation}
\end{figure*}

\section{Observable effects}
\label{sec:effects}

Evidence for a magnetic field is generally sought via Faraday
rotation, but this works only if a source has some linearly polarized
emission.  Below, we show that plasma lensing can provide strong
constraints also for unpolarized sources.  Furthermore, we review the
constraints possible from regular emission using Faraday delay and
generalized Faraday rotation.


\subsection{Birefringence in a plasma lens}
\label{sec:theory}
For a magnetic field of 20\,G, one expects significant birefringence in plasma lensing,
as well as induced circular polarization on unpolarized sources. Below, we show that
birefringence can be identified by looking at the spectra of lensed events in
left and right circular flux separately.
For a uniform magnetic field, we find that spectra should have frequency offsets,
while for a varying magnetic field, we find that the two polarizations should have
different magnifications.  We illustrate both effects in Figure~\ref{fig:simulation}.

\subsubsection{Lensing in a magnetoactive cold plasma}
When the cyclotron frequency is well below the observation frequency,
the natural mode of the plasma will be circularly polarized,
leading to different refractive indices for the two circular polarizations,
\begin{equation}
  n_{L,R}=\sqrt{1-\frac{f_p^2}{f(f\mp f_{B_\parallel})}}
  \simeq 1-\frac{1}{2}\,\frac{f_p^2}{f^2}\left(1\pm\frac{f_{B_\parallel}}{f}\right)
\label{eq:nlr}
\end{equation}
where $L$, $R$ indicate the two circular polarizations,
$f_p=\sqrt{n_e e^2/m_e}=1.6\,\mr{kHz/cm^{3/2}}\,n_e^{1/2}$ is the plasma frequency,
$f$ is the observation frequency,
$f_{B_\parallel}=qB\cos\alpha/2\pi m_e=2.8\,\mr{MHz/G}\,B_\parallel$
is the cyclotron frequency of the parallel magnetic field (with
$\alpha$ is the angle between $\vec{B}$ and the line of sight), and
the approximate equality holds if $f_{B_\parallel}\ll f$ and $f_p\ll f$.

For PSR B1957+20, as is often the case, the plasma and cyclotron
frequencies are much smaller than the observing frequency:
the typical excess DM is of order $10^{-4}\,\DMunit$, which, combined
with a size of order $r_\mr{if}\simeq0.4\,\sr$, implies
$n_e\sim10^{4}{\rm\,cm^{-3}}$ and thus $f_p\sim1\,$MHz, much smaller
than our observing frequency of 330\,MHz. Similarly, for a magnetic
field strength of $\sim\!20$\,G, $f_{B_\parallel}\leq f_B \sim60$\,MHz, which
again is well below 330\,MHz.

For this situation, it is useful to define non-magnetic and magnetic
breaking index differences $\Delta n_p$ and $\Delta n_B$, such that
$n=1-\Delta n_p \mp \Delta n_B$.  With these, after passing through
the plasma, the phase of the electromagnetic wave observed on Earth
can be written as a simple integral along the line of sight,
\begin{align}
    \Phi(\vec{x})
    &=\frac{2\pi}{\lambda}\int (1-\Delta n_p \mp \Delta n_B) dl \\
    &=\Phi_\mr{GM}(\vec{x})+\Phi_\mr{DM}(\vec{x}) \pm \Phi_B(\vec{x})
\label{eq:Phi}
\end{align}
where $\vec{x}=(x,y)$ represents position on the lens plane,
$\Phi_\mr{GM}$ the geometric contribution to the phase,
$\Phi_\mr{DM}$ the dispersive contribution, and
$\Phi_B$ the magnetic contribution.

A pulse will be magnified when the wavefront from a coherent region
is flat, ie. if the extra phase from DM and the magnetic field cancels the
geometric phase, i.e., when
$\Delta\Phi_\mr{GM}+\Delta\Phi_\mr{DM} \pm \Delta\Phi_B\simeq0$, with
the three phase differences given by,
\begin{align}
    \Delta\Phi_\mr{GM}&=\pi\left(\frac{|\vec{x}|}{R_\mr{Fr}}\right)^2\,\\
    \Delta\Phi_\mr{DM}&=2\pi k_\mr{DM}\Delta\mr{DM}(\vec{x})/f\,\\
    \Delta\Phi_B&=\Delta\mr{RM}(\vec{x})\lambda^2
\end{align}
where $R_\mr{Fr}=\sqrt{\lambda a}$ is the Fresnel scale,
with $a$ the distance between pulsar and the plasma lens
and $\lambda\equiv c/f$ the observing wavelength,
$\mr{D\!M}=\int n_e\,dl$ the dispersion measure,
$k_\mr{DM}=e^2/2\pi m_e c=4149{\rm\,s\,MHz^2\,cm^3/pc}$ the dispersion constant,
and $\mr{R\!M}=e^3/2\pi m_e^2 c^4\,\int n_e\Delta B_\parallel\,dl$
the rotation measure.
(Note that, in contrast to \citet{18Main}, our phases
are in radians, not cycles.)

In the above, in principle magnetic fields anywhere along the line of sight
could influence the results.  The strong lensing events we study, however,
last only tens of miliseconds,
while scintillation from the interstellar medium (ISM)
changes on timescales of minutes \citep{17Main}.  Hence, in our measurements of
the magnetic field in the following sections, we can consider the
contribution from the interstellar medium constant, and we probe
effects due to the magnetic field at the pulsar-companion interface.

\subsubsection{Uniform magnetic field}
\label{sec:Bconst}
In the presence of a uniform magnetic field,
$\Delta\Phi_B=\Delta\Phi_\mr{DM}f_{B_\parallel}/f\propto \Delta\mr{DM}\,\mr{B}_\parallel$,
and both the variations of $\Delta\Phi_B$
and $\Delta\Phi_\mr{DM}$ are completely determined by the shape of $\Delta\mr{DM}$.
Therefore, for lensing, the DM has to vary on the same length scale as $\Phi_\mr{GM}$,
i.e., the Fresnel scale,
causing opposite phase delays. Furthermore, in order to avoid cancellation, no variations
should be present on smaller spatial scales.

The parameter governing birefringence is $B_\parallel$.  For a uniform magnetic field,
the cyclotron frequency $f_{B_\parallel}$ will be constant, and
the birefringence will be equivalent to a frequency shift. This can be seen
by considering the refractive index as some small offset frequency $\Delta f$.
To first order,
\begin{align}
    n_{L,R}(f+\Delta f)=1-\frac{1}{2}\,\frac{f_p^2}{f^2}\left(1\pm\frac{f_{B_\parallel}}{f}
    -\frac{2 \Delta f}{f}\right)
\label{eq:ncir}
\end{align}
Therefore, for the same plasma lens in focus at frequency $f$ without a magnetic field, a non-zero $B_\parallel$ will lead to the left circular polarization being focused at $f-f_{B_\parallel}/2$ and the right circular polarization being focused at $f+f_{B_\parallel}/2$.  Thus, the two polarizations will have spectra that are offset
in frequency by $f_{B_\parallel}$, the cyclotron frequency
corresponding to the part of the magnetic field in the lens that is
directed along the line of sight.

The frequency offset should be present in all images formed in plasma lenses,
but it will be measurable only when the magnification spectra are chromatic across the band.
Since the larger the lens, the more chromatic an event should be,
the most magnified events should be the best candidates for measuring frequency shifts.

\subsubsection{Spatially varying magnetic field}
\label{sec:Bvary}
In the presence of a spatially varying magnetic field, the
magnetically induced phase change results from variations in both DM and $B$,
$\Delta\Phi_B\propto \Delta\mr{DM}\,B_\parallel + \mr{DM}\, \Delta B_\parallel$.
As shown in the previous section, the effect of the first term is
a shift in focal frequency between the two polarizations.

The effect of the second term cannot be described as simply, but
it is possible to constrain its amplitude observationally.
In particular, since the sign of $\Delta\Phi_B$ is opposite for the two polarizations,
it is not possible for the wavefront of both polarizations to
be flat at the same time:
if light in one polarization is perfectly lensed by a given region,
then the magnification in the other polarization will be
modified by $|\langle \exp(i\,2\Delta \Phi_B) \rangle|$, where the
average is taken over the lens. For Gaussian fluctuations in $\Delta\Phi_B$,
the reduction factor equals $\exp(-2\langle \Delta\Phi_B^2\rangle)$,
which leads to the intuitive result that light in both polarizations
cannot be magnified similarly at the same time if
$\langle \Delta\Phi_B^2\rangle^{1/2} \gtrsim 1\,$rad.

The constraint on the variations in magnetic field, $\Delta B_\parallel$,
will depend on the excess DM of the lensing region.
For PSR~B1957+20, at the orbital phases
where most magnified pulses occur, $\Delta\mr{DM}\simeq10^{-4}$ to
$10^{-3}\,\DMunit$.  For $\mr{DM}=10^{-4}\,\DMunit$, the constraint that
$\langle\Delta\Phi_B^2\rangle^{1/2}\lesssim1\,$rad implies
$\langle\Delta\Bpara^2\rangle^{1/2}\lesssim10\,\mr{mG}$, i.e., if there were
fluctuations in the magnetic field in excess of $\sim\!10\,$mG on scales of
order the Fresnel scale or smaller, one would expect that the magnification
would be strongly dependent on circular polarization, and thus that
lensing events would be circularly polarized.


\subsection{Faraday delay}
\label{sec:Fdelay}
For a linearly polarized source, magnetized plasma can be easily
detected, either from variations in rotation measure or, if the
variations are large, from depolarization.
In the absence of linearly polarized emission, like for
PSR~B1957+20, one can measure the magnetic field only when the Faraday
rotation is sufficiently large to cause the pulse profiles of the two
circular polarizations to be offset in time.  This effect is known as
\emph{Faraday delay} \citep{90Fruchter}, and equals the difference in
group delay $\tau_F$ between the two circular polarizations,
\begin{equation}
    \tau_F=\frac{4f_{B_\parallel}}{f} \, \tau_p,
\end{equation}
where $\tau_p$ is the mean excess dispersive delay of the two
polarizations.  For $B_\parallel=20\,$G and $f=330\,$MHz, one
infers $\tau_F\simeq0.7\tau_p$.  Below, we will compare Faraday delay
close to and far away from the eclipse to obtain an independent
constraint on $B_\parallel$.

\begin{figure}
  \includegraphics[width=\hsize]{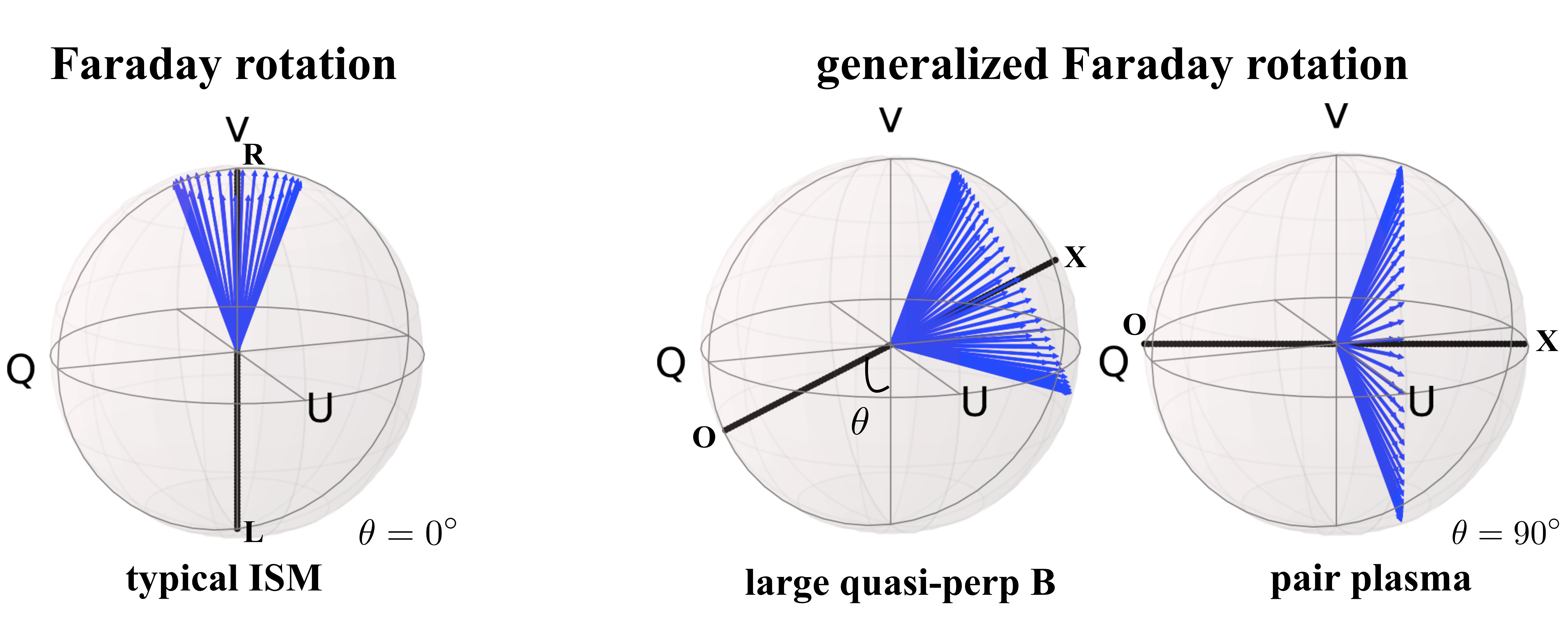}
  \caption{Poincar\'{e} spheres showing rotation of the Stokes
    parameters of an incoming wave (blue arrows) around the natural
    axis (black bar) of the plasma for three situations, a weak or quasi-parallel
    magnetic field leading to a natural axis along the polar axis
    (left), a strong, nearly perpendicular magnetic field giving a
    large angle $\theta$ from the polar axis (middle), and a purely
    perpendicular field with an axis in the equatorial plane (right),
    as appropriate for an electron-positron pair plasma (where the
    parallel terms cancel).  In Faraday rotation (left), the linearly
    polarized emission is rotated and may become depolarized, but
    circularly polarized emission is not affected.  For the
    generalized Faraday rotation, linear and circular polarized
    emission are mixed (and may be depolarized); only the component
    along the natural axis will remain.
\label{fig:faraday}}
\end{figure}

\subsection{Generalized Faraday rotation}
\label{sec:GFR}
As noted in Section~\ref{sec:motivation}, previous observations found
little evidence for parallel magnetic fields in PSR B1957+20, which
might imply that the magnetic field is oriented perpendicular to the
line of sight, or that the excess dispersion is due to a pair plasma.
In either case, there are second-order effects of the magnetic field
which, given that the expected cyclotron frequency of $\sim\!60\,$MHz
is not that far below our observing frequency of $330\,$MHz, might be
detectable.

The effects can be most easily understood by first considering the
natural modes of an electromagnetic wave through a magnetized plasma,
which propagate independently with different phase velocities.  These
typically correspond to the two circular polarizations of the wave,
and the magnetic field induces Faraday rotation.  For a strong
magnetic field perpendicular to the line of sight, however, the
natural modes are two linear polarizations, `o' and `x', for which the
electric vectors are parallel and perpendicular to the magnetic field,
respectively, and one expects a rotation between circular and linear
polarization.

For the general case, the natural modes are elliptical, and one has
\emph{generalized Faraday rotation} \citep{98Kennett}.  One can
visualize the rotation by tracking the change of Stokes parameters
against frequency on a Poincar\'{e} sphere, as we do in
Figure~\ref{fig:faraday}.  Here, the black axis shows the natural axis of
the plasma, which points towards the pole if the natural modes are
circular, somewhere along the equator if the natural modes are linear,
and at some intermediate angle $\theta$ for the general case.  For
incoming waves of fixed polarization but different frequencies,
passing through a magnetized region will cause the Stokes parameters
to be rotated by different amounts, thus leaving them in different
directions around the natural axis of the plasma.

The basis of the natural modes can be quantified by a parameter \citep{94Thompson},
\begin{equation}
  x \equiv 2\left(\frac{f}{f_B}\right)\cos\alpha.
  \label{eq:x}
\end{equation}
where $\alpha$ is the angle between $\vec{B}$ and the line of sight.
For typical conditions, with $f_B\ll f$ and some random $\alpha$, one
has $x\gg1$, resulting in two circular natural modes.  However, $x$
can be small if $\alpha$ is sufficiently close to $90^\circ$, and will
be zero if the dispersing material consists of a pair plasma.

In terms of $x$, for the general case, the refractive indices for the
two natural modes will be,
be \citep{94Thompson},
\begin{equation}
  n^2_\mr{1,2}=1-\frac{f_p^2}{f^2}
    +\frac{f_p^2\,f_B^2}{2f^4}
    \left(1\mp\left(1+x^2\right)^\frac{1}{2}\right).
\label{eq:nli}
\end{equation}
Here, for large $x$, one recovers Eq.~\ref{eq:nlr}, in which the
magnetic term scales linearly with $f_B/f$, while for small $x$
it scales quadratically (and is present also for a pair plasma).

For $x\lesssim1$, the angle $\theta$ between the natural axis of the
plasma and the polar axis of the Poincar\'e sphere is given by,
\begin{equation}
  \cot \theta\simeq\frac{4x}{4-x^2},
  \label{eq:theta}
\end{equation}
and as $x\to0$ the phase difference between the two natural modes
after passing through the plasma becomes,
\begin{align}
    \Delta\Phi_\mr{x,o}&=\frac{2\pi}{\lambda}\int\left(n_\mr{x}-n_\mr{o}\right)\,dl \\
     &\approx \frac{2\pi}{c}\int\frac{f_p^2 f_B^2}{2\,f^3} \,dl
     =\Delta\Phi_\mr{DM}\frac{\langle f_{B_\perp}^2\rangle}{f^2},
\label{eq:phase}
\end{align}
where $\langle f_{B_\perp}^2\rangle$ is an electron-density weighted
average.

Note that the phase difference varies as $1/f^3$, i.e., faster than
the $1/f^2$ for Faraday rotation.  Without correcting this rotation,
the polarized fraction perpendicular to the natural axis will be
averaged down almost linearly with the number of turns, leaving only
the projection of the initial Stokes parameters onto the natural axis.
Inserting numbers for PSR~B1957+20, of $\mr{DM}=10^{-4}\,\DMunit$ and
$f_{B_\perp}=60\,$MHz, and $f=330\,$MHz, one finds
$\Delta\Phi_\mr{DM}\simeq7900\,$rad,
$\Delta\Phi_\mr{x,o}\simeq260\,$rad.  For a band with a width of
48\,MHz centered on 330\,MHz, the rotation varies by about 120\,rad
over the band.  Hence, the effect should be significant.


\begin{figure*}
  \includegraphics[width=\textwidth]{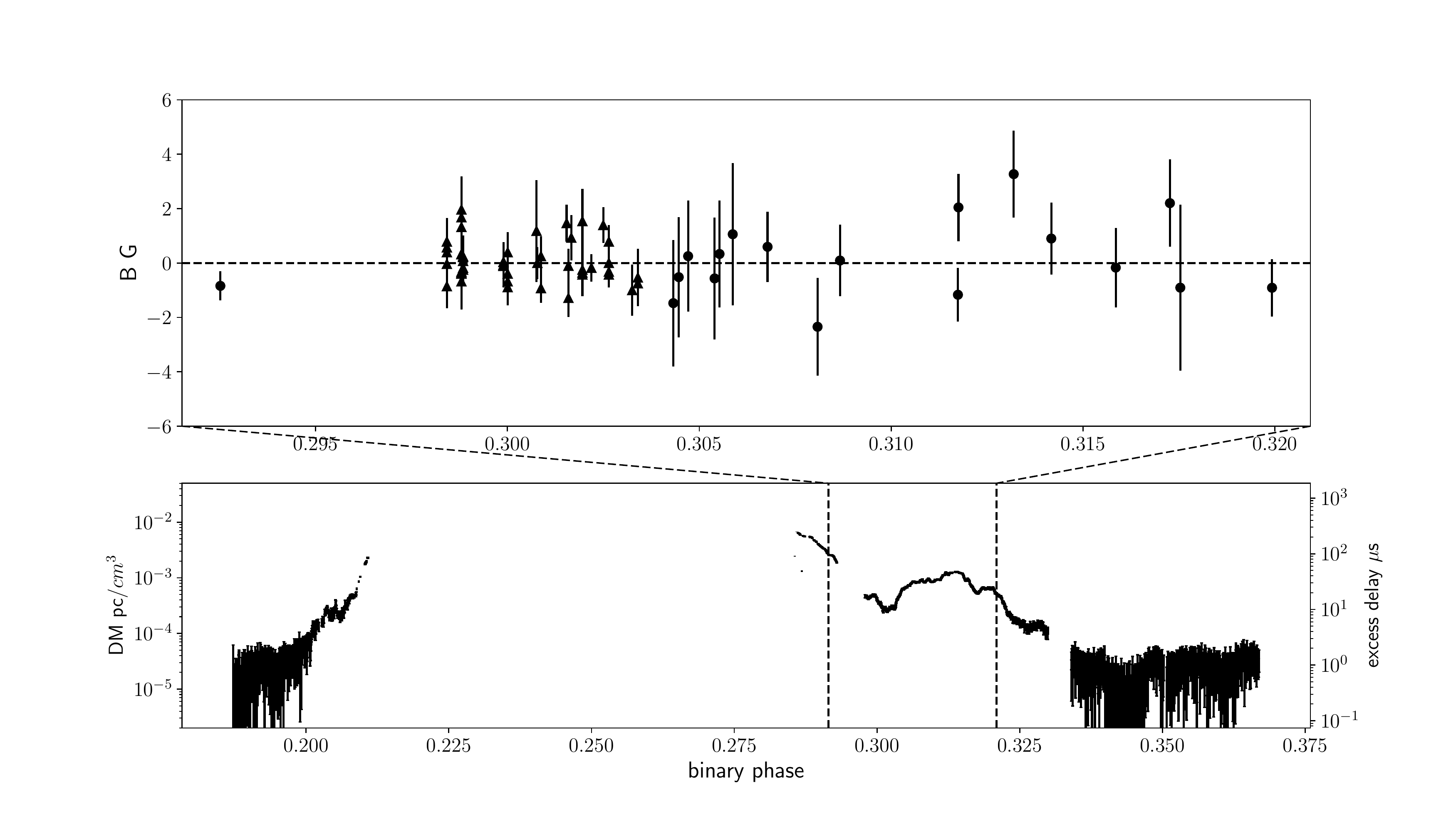}
  \caption{Dispersion and magnetic fields around one eclipse.
    {\em Bottom:\/} excess dispersion measure inferred from excess
    delay in arrival times (right axis) of pulse profiles integrated
    in 2\,s bins.  Our single pulse measurements of parallel magnetic
    fields are done in the section of relatively large excess
    dispersion indicated by the dashed lines.
    {\em Top:\/} Limits on the magnetic field parallel to the line of
    sight, both from frequency offsets between left and right circular
    polarized spectra for pulses magnified by plasma lenses
    (triangles; see Fig.~\ref{fig:freq}), and from Faraday delay
    between left and right circular giant pulse profiles (circles; see Fig.~\ref{fig:arrival}).}
  \label{fig:full}
\end{figure*}

\section{Constraints on the magnetic field in PSR B1957+20}
\label{sec:observations}

We derive constraints on both the parallel and perpendicular magnetic
fields in the regions near eclipse in PSR B1957+20 using 9.5 hours of
baseband data covering 311.25--359.25 MHz.  We took these data in four
2.4 hr sessions on 2014 June 13 to 16 with the 305-m William E. Gordon
Telescope at the Arecibo observatory.
A more detailed description of the
data is given in \citet{17Main}; the method used
to infer excess dispersion is described in \citet{18Main}.

For most of our analysis here, we focus on a 16~min segment
during eclipse egress in which excess dispersion exceeds
$\sim\!10^{-4}\,$\DMunit.  We also focus on the main pulse
component because it is only affected marginally by mode switching
\citep{18Mahajan}.  In this region, we detect 268 pulses which, in the main pulse phase, are at
least 10 times stronger than the
average pulse far away from eclipse.  We classify each pulse based on
its profile (see Appendix \ref{sec:distinguish}), and find that 153
are regular pulses magnified by plasma lenses and 41 are giant pulses
(i.e., intrinsically bright, with unknown magnification).  The
remaining 74 pulses have ambiguous classification and are ignored in
our further analysis.

\begin{figure}
  \includegraphics[width=\columnwidth]{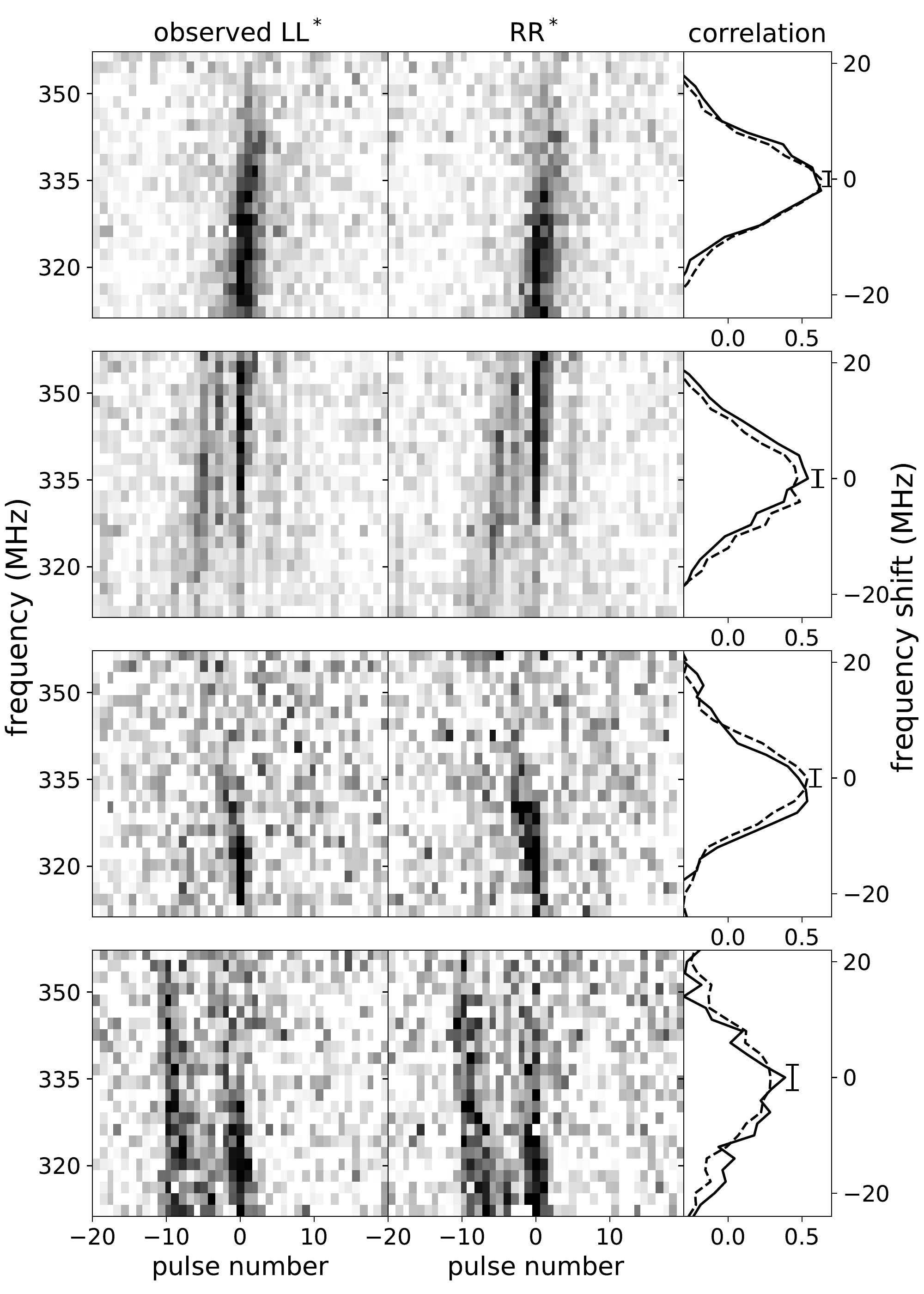}
  \caption{Spectra of four main pulse lensing events in left and right
    circular polarization.  The images show the flux in 2\,MHz bins
    for consecutive pulses, and the side panels show the cross
    polarization frequency correlation of the brightest pulse (number
    0).  The similarity of the spectra of the two polarizations sets
    strong limits on the strength of any magnetic field parallel to
    the line of sight.}
  \label{fig:freq}
\end{figure}

\subsection{Constraints from lensed pulses}
We first constrain the parallel magnetic field using the spectra of
the pulses magnified by plasma lenses in left and right circular
polarization.  Four of these pairs of spectra are shown in
Figure~\ref{fig:freq}.  Here, we follow \citet{18Main} and divide the
spectra by average spectra from the surrounding 30\,s to remove
effects of interstellar scintillation (which is constant on these
timescales, and identical for the two polarizations).  From the
figure, it is clear that the spectra of the two polarizations do not
differ significantly; below, we discuss the resulting constraints on
a uniform or small-scale magnetic field.

\subsubsection{Medium and large-scale parallel magnetic fields}
\label{sec:seeBconst}
The presence of a uniform magnetic field in a lensing region will, as
shown in section \ref{sec:Bconst}, generate a difference in focal
frequency between left and right polarizations.  To obtain the
strongest constraint, we select the pulses with the strongest
chromaticity across our band and cross-correlate the left and right
polarized flux.  Next, we determine the frequency offset by fitting a
Gaussian to the cross correlation function, and an associated
uncertainty from simulations, in which for each selected pulse we take
the smoothed polarization-averaged frequency profile, add independent
sets of random noises to simulate the two polarizations, and then find
frequency offsets in the cross-correlation functions as for the real
data.
We take as the uncertainty for a given pulse
the $68\%$ confidence interval in the offsets measured in 200 such
simulations.

Inspecting the spectra, we find that the pulses are very similarly
magnified in both polarizations (see Fig.~\ref{fig:freq}): the auto
and cross correlations of their spectra are equally strong and no
statistically significant offset in the cross correlation is found.
The $B_\parallel$ derived from the spectra
are shown as triangles in Figure~\ref{fig:full}.  Each
magnified event constrains $B_\parallel$ on the scales of the lensing
region to be smaller than 1 to 3~G.  Averaging 51 events, we
constrain any large-scale uniform magnetic field to
$\overline{B}_\parallel = 0.02\pm0.09$~G,
with a reduced $\chi^2=0.9$ ---
this is consistent with zero magnetic field.
\subsubsection{Small-scale parallel magnetic field variations}
Variations in magnetic field strength within the lensing regions will
make it impossible for a lens to focus left and right circular the same
way, and in Section \ref{sec:Bvary} we found that even small changes
in $B_\parallel$ would result in only a single polarization being
magnified (see Figure \ref{fig:simulation}).  For all the detected
lensing events,both polarizations are magnified, by amounts that are
the same to within the uncertainties.  Quantitatively, we infer that
within those lenses, $\left\langle\Delta
B_\parallel^2\right\rangle^{1/2}<10\,\mr{mG}$.

\subsection{Constraints from giant pulses}
\label{sec:verification}
PSR B1957+20 shows short, intrinsically bright giant pulses
\citep{06Knight,17Main}, which are strongly polarized and can be used
to provide independent measurements of the magnetic field.

The polarization properties of giant pulses vary from pulse to pulse.
For magnetic field measurements, the best ones are those which are
strongly linearly polarized and thus allow one to measure Faraday
rotation.  During the 7.2\,hr observation away from eclipse, we found
28 suitable giant pulses, but unfortunately none of the giant pulses
in the $\sim\!0.5$\,hr near eclipse were similarly suitable.
Therefore, we instead attempt to measure the Faraday delay of those
giant pulses, which gives a complementary constraint to that derived
by \cite{90Fruchter} from profiles integrated over 10--60\,s, in that
it is sensitive also to magnetic fields that vary on much smaller
scales.  Furthermore, we use one bright giant pulse with strong
circular polarization to constrain the perpendicular magnetic field.

\begin{figure}
  \includegraphics[width=\columnwidth]{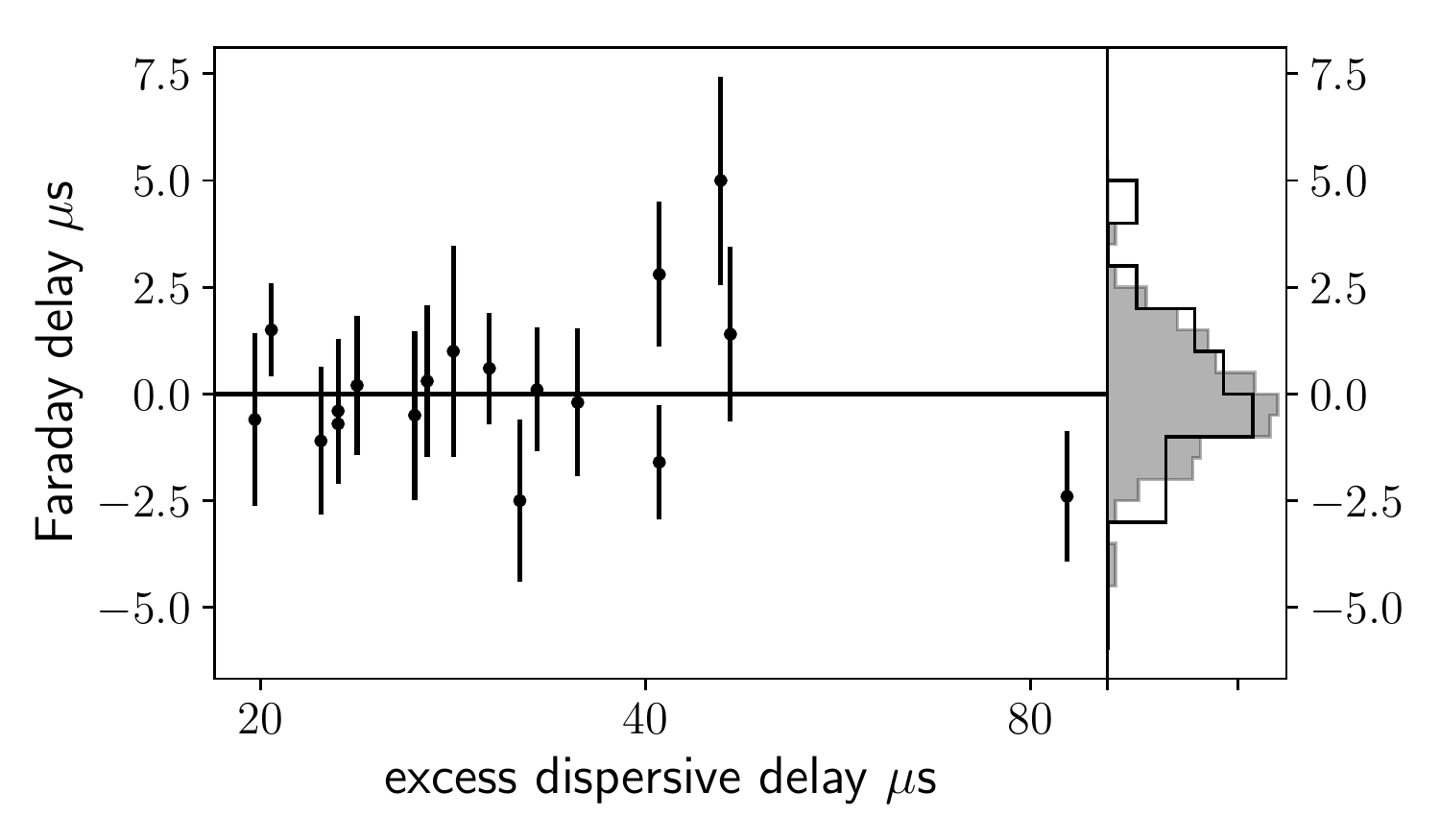}
  \caption{Faraday delay measured for individual giant pulses.
    {\em Left:\/} Delays $\tau_F$ for individual giant pulses near
    eclipse, as a function of the excess dispersive delay near them.
    No significant detections are made.
    {\em Right:\/} Delay histograms for giant pulses near (line) and
    away (grey, filled) from eclipse, showing that the distributions
    are the same within the uncertainties. }
  \label{fig:arrival}
\end{figure}

\subsubsection{Medium and large-scale parallel magnetic fields}

We measure the Faraday delay $\tau_F$ by cross-correlating giant pulse
profiles in left and right circular polarizations, and fitting a
Gaussian to the cross-correlation function.
We perform the same bootstrap simulation as described in section \ref{sec:seeBconst}
to obtain 1$\sigma$ error for the Faraday delay.
We compare the simulated errors for GPs away from eclipses, where $\tau_F$ is expected to be 0,
with the actual scattering of measured $\tau_F$ from the same group.
The average simulated error is $\sigma=1.5\, \mu$s, and the scattering of measured $\tau_F$ is $\sigma=1.3\,\mu$s,
which is consistent.
For each GP, we measure the dispersive delay $\tau_p$ near it by
fitting the regular pulse profile integrated over 2\,s to the average
pulse profile away from eclipse (enough to ensure the uncertainty, of
$\sim\!0.5\,\mu$s, is small compared that for the time offset).

In Figure~\ref{fig:arrival}, we show the measured Faraday delay for
all the GPs with excess dispersive delay $\tau_p > 20\,\mu s$.  As can
be seen, no significant detections are made for any giant pulse, and
indeed the distribution of the measured $\tau_F$ near eclipse is the
same as that for the ones measured away from eclipse.  The
corresponding limits on the parallel magnetic field are
$B_\parallel\lesssim 5$\,G (see circles in the top panel of
Figure~\ref{fig:full}).  One also sees that the measured Faraday delay
is not increasing with dispersive delay, as would be expected for a
large-scale uniform magnetic field; averaging the constraints yields
$\overline{B}_\parallel=-0.2\pm0.3$\,G. The data are thus
  consistent with $B_\parallel=0\,$G, with a reduced $\chi^2=0.9$ for N=18.

\begin{figure}
  \includegraphics[width=0.9\hsize]{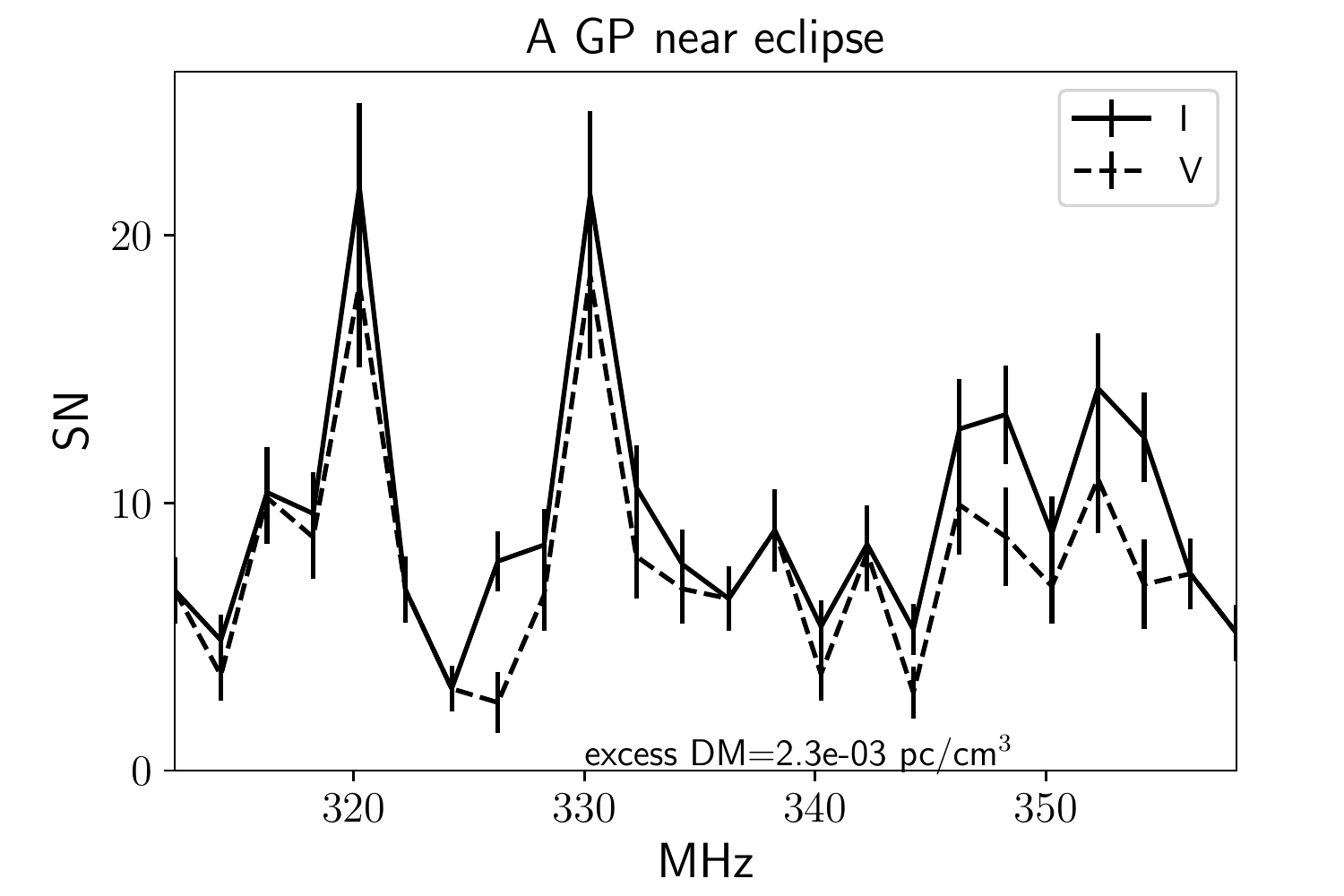}
  \caption{Spectrum of a giant pulse in a region of high excess
    dispersion measure, of $\sim\!10^{-3}\,\DMunit$.  The pulse is
    strongly circularly polarized across the band, without any
    evidence of dips or sign changes due to generalized Faraday
    rotation.}
    \label{fig:V_GP}
\end{figure}

\subsubsection{Perpendicular magnetic field}
\label{sec:V_gp}
We detect one bright giant pulse near eclipse, in the region with
excess dispersion of $10^{-3}$\DMunit, which is strongly circularly
polarized.  As can be seen in Figure~\ref{fig:V_GP}, the polarization
is strong across the band, i.e., it is not depolarized by rapid
generalized Faraday rotation around a natural axis close to the
Poincar\'{e} sphere equator, as would be given the expected
$\sim\!200$ turns expected across the band for a 20\,G field (see
Sec.~\ref{sec:GFR}).  Indeed, the circular polarization does not
change sign across the band, suggesting that the induced rotation
changes by less than a cycle across the band,
which strongly against a quasi-perpendicular magnetic field
(for detailed discussion, see section \ref{sec:bperp}).

\begin{figure}
  \includegraphics[width=\hsize]{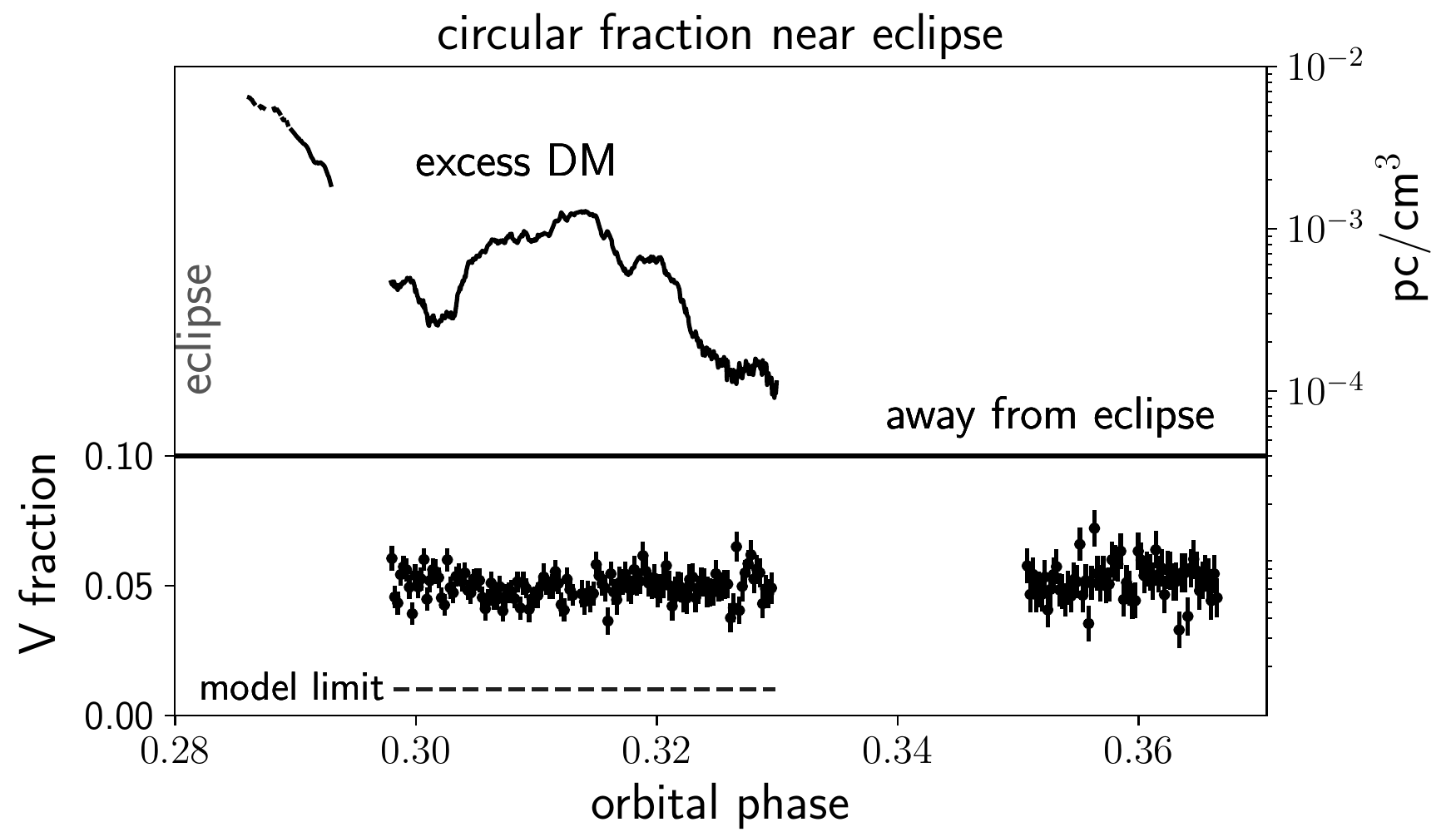}
  \vspace{-0.3cm}
  \caption{Circularly polarized fraction of the emission of the main
    pulse, averaged in 10\,s bins.  The fraction at large excess
    dispersion measure is consistent with that seen away from eclipse,
    unlike what would be expected if the eclipse region had a strong
    perpendicular magnetic field.}
    \label{fig:V_frac}
\end{figure}

\subsection{Constraints from the average pulse profile}
\label{sec:prof_V}

The average pulse profile has a small circularly polarized component.
If there were a perpendicular magnetic field that was much stronger
than the parallel one, one would expect this to be reduced by
generalized Faraday rotation.  To look for this, we measured the
cirularly polarized fraction for pulse profiles integrated over 10\,s.
We find that both near and away from eclipse, the circularly polarized
fraction is $5\pm 1\%$, without any correlation with excess dispersion
measure (see Fig.~\ref{fig:V_frac}).  This suggests no more than about
most a turn across our band due to generalized Faraday rotation,
which rules out the hypothesis of a quasi-perpendicular magnetic field (see also Sec.~\ref{sec:bperp} below).

\section{Implications}
\label{sec:alternatives}
We described in Section \ref{sec:motivation} the theoretical arguments
for expecting a strong, $\gtrsim\!20\,G$ magnetic field near the eclipse
region, and gave three possible reasons for why, nevertheless,
\cite{90Fruchter} might have failed to detect such fields using
Faraday delay in integrated pulse profiles.  We discuss how our new
constrains affect these scenarios.

\subsection{Small-scale variations in magnetic field}
The magnetic field could be as strong as expected, but organized in
small loops, with radii of order $cP/2\approx240$\,km.  In Section
\ref{sec:seeBconst}, however, we found that even for single magnified
pulses, no effect of $B_\parallel$ is seen, and hence we exclude
fields on spatial scales larger than those of the lenses, i.e., a few
times the Fresnel scale.  We also found that the field cannot vary
much on scales or order of or smaller than the Fresnel scale, with
$\sigma_{B_\parallel}=\langle\Delta B_\parallel^2\rangle^{1/2}<10\,\mr{mG}$
in the lensing region.

The reconnection loops can provide a net $B_\parallel$ of exactly zero
spatially along the loop, but they will inevitably yield magnetic
variance.  Given previous constraints, the only way to obtain a low
$B_\parallel$ was to average it down along the line of sight by having
random orientations of the magnetic fields in different loops.  Our
new constraints on the magnetic field in small spatial structures
imply that this cancellation has to be better than 10\,mG for each
line of sight.  To average down $B_\parallel$ from $\sim\!10$\,G to
$\lesssim\!10$\,mG, one would have to average over $\gtrsim\!10^{6}$
reconnection loops, well beyond the maximum of $\sim a/(cP/2)\sim
10^{3}$ reconnection loops that can reasonably be expected.  The
scenario of a highly varying large magnetic field thus becomes very
unlikely at this point.


\subsection{Perpendicular magnetic field}
\label{sec:bperp}
The low observed $B_\parallel$ could still be consistent with a
strong, $\gtrsim\!20\,$G overall magnetic field if the field were nearly
perpendicular to the LOS.
In this case, the magnetic field would necessarily have a large scale,
and by averaging our constraints given in Section \ref{sec:seeBconst},
we found $B_\parallel=0.03\pm0.09\,$G with reduced $\chi^2=1.0$ for
all 69 measurements.  To accommodate a 20\,G
magnetic field with parallel fraction smaller than $0.2\,$G, the angle $\alpha$ between $\vec{B}$ and the line of
sight would have to be within $0.6^\circ$ of perpendicular.  In this
case, from Equations \ref{eq:x} and \ref{eq:theta}, $x\lesssim0.3$ and
$\theta\gtrsim 70^\circ$, i.e., the natural mode of the plasma is no
longer circular but strongly elliptical.

Given excess dispersions between $10^{-4}$ and $10^{-3}\,\DMunit$ in
the lensing region, one expects the Stokes parameters to rotate around
the axis of the natural modes, with a number of cycles that differs by
between 20 and 200 cycles across our 48\,MHz bandwidth (see
Sec.~\ref{sec:GFR}).  This is enough to average down the polarization
fraction perpendicular to the natural mode by at least an order of
magnitude.  For $\theta\gtrsim70^\circ$, given that the circular and
linear fraction away from eclipse are $V_\mr{frac}\sim\!5$\% and $L_\mr{frac}<\!2\%$,
respectively, one would thus expect an observed circular fraction near eclipse of
$V_\mr{frac}^\prime<(V_\mr{frac}\cos(\theta)+L_\mr{frac}\sin(\theta))\cos(\theta)=1\%$.
Since this is inconsistent with the measurements shown in
Figure~\ref{fig:V_frac}, we conclude that a large quasi-perpendicular
magnetic field is not present in the interface.

\subsection{Pair plasma}
If the excess dispersion measure is due not just to electrons, but
reflects an electron-positron pair plasma, then the effects of a
parallel magnetic field will cancel, because electrons and positrons
cause Faraday rotation in opposite directions.  A pair plasma could
not, however, conceal a perpendicular magnetic field, because the gyro
motions of electrons and positrons along such a field project in the
same way on the linear polarization of light, i.e., for both the phase
velocity of the `o' mode is that of an unmagnetized plasma, while that
of the `x' mode is not as strongly affected.  Hence, the natural modes
for a pair plasma are two linear polarizations, which are identical to
what one would find if one had just electrons in a completely
perpendicular magnetic field \citep{97Melrose}.

Given the above, for a magnetized pair plasma one expects the Stokes
parameters of the incoming wave to rotate around a natural axis
located in the equator of the Poincar\'{e} sphere \ref{fig:faraday}.
From Equation \ref{eq:phase}, assuming a 20\,G magnetic field at
30$^\circ$ from the line of sight (i.e., $B_\perp=10\,$G), one expects
rotations by 10 to 100 cycles across the band, again strongly
suppressing the circularly polarized fraction, in contrast to what we
observe.  For a pair plasma, the only solution would seem to assume a
largely {\em parallel} magnetic field, such that
$B_\perp\lesssim0.1\,$G.  This seems contrived.

\section{Ramifications}
\label{sec:conclusion}
In this paper, we have demonstrated that birefringence in plasma
lenses can be used to measure the parallel magnetic field
$B_\parallel$ and to constrain small-scale variations in that field.
We attempted to do so for the Black Widow Pulsar B1957+20, where $B
\gtrsim 20$~G is expected at the interface between the pulsar wind and
its companion, but found only upper limits, which are inconsistent
with the presence of such a strong field, even if it were to vary
rapidly in orientation.

We also found that other scenarios to conceal a large magnetic field,
such as placing it nearly perpendicular to the line of sight, or
assuming that the excess dispersion is caused by a pair plasma,
seem excluded as well, as they would predict changes in the fraction of
circular polarization that are not observed.  Given these new
constraints, the long eclipse and the large stand-off distance of the
companion wind are very puzzling.

Birefringence in plasma lensing could be applied to other systems in
which plasma lensing has been seen or suspected.  For instance, the
binary pulsar PSR J1748-2446A also shows magnified events near eclipse
\citep{11Bilous}.  It would also be interesting to look for further
systems, in particular ones such as PSR J1748-2446 and PSR J2256-1024
in which changes in RM and/or depolarization are observed near
eclipse \cite{18You, 18Crowter}.
Furthermore, Fast Radio Bursts might be good targets, as they have been
proposed to be lensed by plasma in their local galaxies \citep{17Cordes}.
Indeed, the spectra
of the only repeating FRB so far, FRB 121102, look similar to the
lensing spectra of B1957+20 \citep{16Spitler, 18Farah,18Main}, and
their strong chromaticity should make it relatively easy to detect the
frequency offsets resulting from birefringence.  Furthermore, a dynamic
magneto-ionic environment has been inferred for FRB 121102
\citep{18Michilli} and it is fairly clear that other FRBs are observed
to pass through dense magnetized plasma in their host galaxy (e.g.,
FRB 110523; \citealt{15Masui}).  If birefringence is found, it would
provide direct proof that FRBs are affected by lensing, and would give
a local measurement of the field strength (unlike a Rotation Measure,
which is integrated along the line of sight).

\section*{Acknowledgements}
We greatly appreciated discussions with and help from Chris Thompson,
Almog Yalinewich, and Daniel Baker, as well as comments from Daniele
Michilli and Anna Bilous.  We made use of NASA'a Astrophysics Data
System and SOSCIP Consortium's Blue Gene/Q computing platform.  The
Arecibo Observatory is a facility of the National Science Foundation
(NSF) operated by SRI International in alliance with the Universities
Space Research Association (USRA) and UMET under a cooperative
agreement.  This research made use of Astropy, a community-developed
core Python package for Astronomy \citep{Astropy} and Baseband
(\url{http://baseband.readthedocs.io})




\bibliographystyle{mnras}
\bibliography{eclipse_B} 

\begin{thebibliography}{}
\makeatletter
\relax
\def\mn@urlcharsother{\let\do\@makeother \do\$\do\&\do\#\do\^\do\_\do\%\do\~}
\def\mn@doi{\begingroup\mn@urlcharsother \@ifnextchar [ {\mn@doi@}
  {\mn@doi@[]}}
\def\mn@doi@[#1]#2{\def\@tempa{#1}\ifx\@tempa\@empty \href
  {http://dx.doi.org/#2} {doi:#2}\else \href {http://dx.doi.org/#2} {#1}\fi
  \endgroup}
\def\mn@eprint#1#2{\mn@eprint@#1:#2::\@nil}
\def\mn@eprint@arXiv#1{\href {http://arxiv.org/abs/#1} {{\tt arXiv:#1}}}
\def\mn@eprint@dblp#1{\href {http://dblp.uni-trier.de/rec/bibtex/#1.xml}
  {dblp:#1}}
\def\mn@eprint@#1:#2:#3:#4\@nil{\def\@tempa {#1}\def\@tempb {#2}\def\@tempc
  {#3}\ifx \@tempc \@empty \let \@tempc \@tempb \let \@tempb \@tempa \fi \ifx
  \@tempb \@empty \def\@tempb {arXiv}\fi \@ifundefined
  {mn@eprint@\@tempb}{\@tempb:\@tempc}{\expandafter \expandafter \csname
  mn@eprint@\@tempb\endcsname \expandafter{\@tempc}}}

\bibitem[\protect\citeauthoryear{{Arzoumanian}, {Fruchter}  \&
  {Taylor}}{{Arzoumanian} et~al.}{1994}]{94Arzoumanian}
{Arzoumanian} Z.,  {Fruchter} A.~S.,   {Taylor} J.~H.,  1994, \mn@doi [\apjl]
  {10.1086/187346}, \href {http://adsabs.harvard.edu/abs/1994ApJ...426L..85A}
  {426, 85}

\bibitem[\protect\citeauthoryear{{Astropy Collaboration} et~al.,}{{Astropy
  Collaboration} et~al.}{2013}]{Astropy}
{Astropy Collaboration} et~al., 2013, \mn@doi [\aap]
  {10.1051/0004-6361/201322068}, \href
  {https://ui.adsabs.harvard.edu/#abs/2013A&A...558A..33A} {558, A33}

\bibitem[\protect\citeauthoryear{{Berdyugina}, {Harrington}, {Kuzmychov},
  {Kuhn}, {Hallinan}, {Kowalski}  \& {Hawley}}{{Berdyugina}
  et~al.}{2017}]{17Berdyugina}
{Berdyugina} S.~V.,  {Harrington} D.~M.,  {Kuzmychov} O.,  {Kuhn} J.~R.,
  {Hallinan} G.,  {Kowalski} A.~F.,   {Hawley} S.~L.,  2017, \mn@doi [\apj]
  {10.3847/1538-4357/aa866b}, \href
  {http://adsabs.harvard.edu/abs/2017ApJ...847...61B} {847, 61}

\bibitem[\protect\citeauthoryear{{Bilous}, {Ransom}  \& {Nice}}{{Bilous}
  et~al.}{2011}]{11Bilous}
{Bilous} A.~V.,  {Ransom} S.~M.,   {Nice} D.~J.,  2011, in {Burgay} M.,
  {D'Amico} N.,  {Esposito} P.,  {Pellizzoni} A.,   {Possenti} A.,  eds,
  American Institute of Physics Conference Series Vol. 1357, American Institute
  of Physics Conference Series. pp 140--141, \mn@doi{10.1063/1.3615100}

\bibitem[\protect\citeauthoryear{{Born} \& {Wolf}}{{Born} \&
  {Wolf}}{1980}]{80Born}
{Born} M.,  {Wolf} E.,  1980, {Principles of Optics Electromagnetic Theory of
  Propagation, Interference and Diffraction of Light}

\bibitem[\protect\citeauthoryear{{Cordes}, {Wasserman}, {Hessels}, {Lazio},
  {Chatterjee}  \& {Wharton}}{{Cordes} et~al.}{2017}]{17Cordes}
{Cordes} J.~M.,  {Wasserman} I.,  {Hessels} J.~W.~T.,  {Lazio} T.~J.~W.,
  {Chatterjee} S.,   {Wharton} R.~S.,  2017, \mn@doi [\apj]
  {10.3847/1538-4357/aa74da}, \href
  {http://adsabs.harvard.edu/abs/2017ApJ...842...35C} {842, 35}

\bibitem[\protect\citeauthoryear{Crowter}{Crowter}{2018}]{18Crowter}
Crowter K.,  2018, PhD thesis, University of British Columbia,
  \mn@doi{http://dx.doi.org/10.14288/1.0371173}, \url
  {https://open.library.ubc.ca/collections/ubctheses/24/items/1.0371173}

\bibitem[\protect\citeauthoryear{{Farah} et~al.,}{{Farah}
  et~al.}{2018}]{18Farah}
{Farah} W.,  et~al., 2018, \mn@doi [\mnras] {10.1093/mnras/sty1122}, \href
  {http://adsabs.harvard.edu/abs/2018MNRAS.tmp.1067F} {}

\bibitem[\protect\citeauthoryear{{Fruchter}, {Stinebring}  \&
  {Taylor}}{{Fruchter} et~al.}{1988}]{88Fruchter}
{Fruchter} A.~S.,  {Stinebring} D.~R.,   {Taylor} J.~H.,  1988, \mn@doi [\nat]
  {10.1038/333237a0}, \href {http://adsabs.harvard.edu/abs/1988Natur.333..237F}
  {333, 237}

\bibitem[\protect\citeauthoryear{{Fruchter} et~al.,}{{Fruchter}
  et~al.}{1990}]{90Fruchter}
{Fruchter} A.~S.,  et~al., 1990, \mn@doi [\apj] {10.1086/168502}, \href
  {http://adsabs.harvard.edu/abs/1990ApJ...351..642F} {351, 642}

\bibitem[\protect\citeauthoryear{{Kaplan}, {Bhalerao}, {van Kerkwijk},
  {Koester}, {Kulkarni}  \& {Stovall}}{{Kaplan} et~al.}{2013}]{13Kaplan}
{Kaplan} D.~L.,  {Bhalerao} V.~B.,  {van Kerkwijk} M.~H.,  {Koester} D.,
  {Kulkarni} S.~R.,   {Stovall} K.,  2013, \mn@doi [\apj]
  {10.1088/0004-637X/765/2/158}, \href
  {http://adsabs.harvard.edu/abs/2013ApJ...765..158K} {765, 158}

\bibitem[\protect\citeauthoryear{{Kennett} \& {Melrose}}{{Kennett} \&
  {Melrose}}{1998}]{98Kennett}
{Kennett} M.,  {Melrose} D.,  1998, \mn@doi [\pasa] {10.1071/AS98211}, \href
  {http://adsabs.harvard.edu/abs/1998PASA...15..211K} {15, 211}

\bibitem[\protect\citeauthoryear{{Knight}, {Bailes}, {Manchester}, {Ord}  \&
  {Jacoby}}{{Knight} et~al.}{2006}]{06Knight}
{Knight} H.~S.,  {Bailes} M.,  {Manchester} R.~N.,  {Ord} S.~M.,   {Jacoby}
  B.~A.,  2006, \mn@doi [\apj] {10.1086/500292}, \href
  {http://adsabs.harvard.edu/abs/2006ApJ...640..941K} {640, 941}

\bibitem[\protect\citeauthoryear{{Mahajan}, {van Kerkwijk}, {Main}  \&
  {Pen}}{{Mahajan} et~al.}{2018}]{18Mahajan}
{Mahajan} N.,  {van Kerkwijk} M.~H.,  {Main} R.,   {Pen} U.-L.,  2018,
  preprint, \href {https://ui.adsabs.harvard.edu/#abs/2018arXiv180701713M} {p.
  arXiv:1807.01713} (\mn@eprint {arXiv} {1807.01713})

\bibitem[\protect\citeauthoryear{{Main}, {van Kerkwijk}, {Pen}, {Mahajan}  \&
  {Vanderlinde}}{{Main} et~al.}{2017}]{17Main}
{Main} R.,  {van Kerkwijk} M.,  {Pen} U.-L.,  {Mahajan} N.,   {Vanderlinde} K.,
   2017, \mn@doi [\apjl] {10.3847/2041-8213/aa6f03}, \href
  {http://adsabs.harvard.edu/abs/2017ApJ...840L..15M} {840, L15}

\bibitem[\protect\citeauthoryear{{Main} et~al.,}{{Main} et~al.}{2018}]{18Main}
{Main} R.,  et~al., 2018, \mn@doi [\nat] {10.1038/s41586-018-0133-z}, \href
  {http://adsabs.harvard.edu/abs/2018Natur.557..522M} {557, 522}

\bibitem[\protect\citeauthoryear{{Masui} et~al.,}{{Masui}
  et~al.}{2015}]{15Masui}
{Masui} K.,  et~al., 2015, \mn@doi [\nat] {10.1038/nature15769}, \href
  {http://adsabs.harvard.edu/abs/2015Natur.528..523M} {528, 523}

\bibitem[\protect\citeauthoryear{{Melrose}}{{Melrose}}{1997}]{97Melrose}
{Melrose} D.~B.,  1997, \mn@doi [Plasma Physics and Controlled Fusion]
  {10.1088/0741-3335/39/5A/010}, \href
  {http://adsabs.harvard.edu/abs/1997PPCF...39...93M} {39, A93}

\bibitem[\protect\citeauthoryear{{Michilli} et~al.,}{{Michilli}
  et~al.}{2018}]{18Michilli}
{Michilli} D.,  et~al., 2018, \mn@doi [\nat] {10.1038/nature25149}, \href
  {http://adsabs.harvard.edu/abs/2018Natur.553..182M} {553, 182}

\bibitem[\protect\citeauthoryear{{Ryba} \& {Taylor}}{{Ryba} \&
  {Taylor}}{1991}]{91Ryba}
{Ryba} M.~F.,  {Taylor} J.~H.,  1991, \mn@doi [\apj] {10.1086/170613}, \href
  {http://adsabs.harvard.edu/abs/1991ApJ...380..557R} {380, 557}

\bibitem[\protect\citeauthoryear{{Schneider}, {Ehlers}  \& {Falco}}{{Schneider}
  et~al.}{1992}]{92Schneider}
{Schneider} P.,  {Ehlers} J.,   {Falco} E.~E.,  1992, {Gravitational Lenses},
  \mn@doi{10.1007/978-3-662-03758-4.
}

\bibitem[\protect\citeauthoryear{{Spitler} et~al.,}{{Spitler}
  et~al.}{2016}]{16Spitler}
{Spitler} L.~G.,  et~al., 2016, \mn@doi [\nat] {10.1038/nature17168}, \href
  {http://adsabs.harvard.edu/abs/2016Natur.531..202S} {531, 202}

\bibitem[\protect\citeauthoryear{{Steiner}, {Gandolfi}, {Fattoyev}  \&
  {Newton}}{{Steiner} et~al.}{2015}]{15Steiner}
{Steiner} A.~W.,  {Gandolfi} S.,  {Fattoyev} F.~J.,   {Newton} W.~G.,  2015,
  \mn@doi [\prc] {10.1103/PhysRevC.91.015804}, \href
  {http://adsabs.harvard.edu/abs/2015PhRvC..91a5804S} {91, 015804}

\bibitem[\protect\citeauthoryear{{Stovall} et~al.,}{{Stovall}
  et~al.}{2014}]{14Stovall}
{Stovall} K.,  et~al., 2014, \mn@doi [\apj] {10.1088/0004-637X/791/1/67}, \href
  {http://adsabs.harvard.edu/abs/2014ApJ...791...67S} {791, 67}

\bibitem[\protect\citeauthoryear{{Thompson}, {Blandford}, {Evans}  \&
  {Phinney}}{{Thompson} et~al.}{1994}]{94Thompson}
{Thompson} C.,  {Blandford} R.~D.,  {Evans} C.~R.,   {Phinney} E.~S.,  1994,
  \mn@doi [\apj] {10.1086/173728}, \href
  {http://adsabs.harvard.edu/abs/1994ApJ...422..304T} {422, 304}

\bibitem[\protect\citeauthoryear{{Tuntsov}, {Walker}, {Koopmans}, {Bannister},
  {Stevens}, {Johnston}, {Reynolds}  \& {Bignall}}{{Tuntsov}
  et~al.}{2016}]{16Tuntsov}
{Tuntsov} A.~V.,  {Walker} M.~A.,  {Koopmans} L.~V.~E.,  {Bannister} K.~W.,
  {Stevens} J.,  {Johnston} S.,  {Reynolds} C.,   {Bignall} H.~E.,  2016,
  \mn@doi [\apj] {10.3847/0004-637X/817/2/176}, \href
  {http://adsabs.harvard.edu/abs/2016ApJ...817..176T} {817, 176}

\bibitem[\protect\citeauthoryear{{You}, {Manchester}, {Coles}, {Hobbs}  \&
  {Shannon}}{{You} et~al.}{2018}]{18You}
{You} X.~P.,  {Manchester} R.~N.,  {Coles} W.~A.,  {Hobbs} G.~B.,   {Shannon}
  R.,  2018, preprint, \href
  {https://ui.adsabs.harvard.edu/#abs/2018arXiv180901309Y} {p.
  arXiv:1809.01309} (\mn@eprint {arXiv} {1809.01309})

\bibitem[\protect\citeauthoryear{{van Kerkwijk}, {Breton}  \& {Kulkarni}}{{van
  Kerkwijk} et~al.}{2011}]{11Marten}
{van Kerkwijk} M.~H.,  {Breton} R.~P.,   {Kulkarni} S.~R.,  2011, \mn@doi
  [\apj] {10.1088/0004-637X/728/2/95}, \href
  {http://adsabs.harvard.edu/abs/2011ApJ...728...95V} {728, 95}

\makeatother
\end{thebibliography}

\appendix

\section{Simulation with birefringence}
\label{sec:sim}
We demonstrate the effects of birefringence in plasma lens with a simple
simulation, in which we assume one-dimensional variations in
dispersion measure,\footnote{For a two-dimensional simulation, the
  dependence of chromaticity on magnification changes, but the effects
  due to magnetic fields change very little, as one can see from the
  derivations in section \ref{sec:theory}.}  and aim to mimic the
conditions found in PSR B1957+20.  We start by generating a randomly
varying dispersion measure phase screen $\Phi_\text{DM}$, ensuring
that at large scales we match the observed power spectrum
$P_\text{DM}(q)$ of excess DM in the egress of the eclipse. Here, we
assume a flow velocity three times the orbital velocity to convert
timescales on which we measure dispersion measure to spatial scales
$1/q$ 
-- an arbitrarily selection within the constraints from
\citet{18Main} to make simulated spectra look like real spectra.
With this choice, and imposing an inner scale cutoff
at $q_\text{in}$ in the power spectrum in order for the strong
magnification to appear, we obtain output spectra that resemble
observed ones.  Specifically, $\Phi_\text{DM}$ is computed as
\begin{align}
  \Phi_\text{DM}(x) = \text{FT}^{-1}\left[
    2\pi(u_q + i v_q)\sqrt{P_\text{DM}(q)}
    \exp\left(-\frac{q^2}{2q_\text{in}^2}\right)
    \right]
\end{align}
where FT is the Fourier transform and $u_q$ and $v_q$ are normally
distributed random variables with unit variance. For Fig \ref{fig:simulation},
we used $1/q_\mr{in} \sim 7600$ km, a relatively large inner scale cutoff
in order to see enough strong lenses in short simulation time, and $P_\mr{DM}(q) \sim 1.3\cdot 10^{-9}
q^{-2.0}$ to measure the observed DM distribution.

At each point of the phase screen, we add a magnetic field
contribution $\Phi_B=\Phi_\mr{DM}(x)f_{B_\parallel}(x)/f$ (with
$f_{B_\parallel}(x)$ reflecting a choice of magnetic field
distribution), and then compute the received electric field using the
Fresnel-Kirchhoff integral \citep{80Born}:
\begin{equation}
E_r (f, x_s) = \frac{e^{-i \pi/4}}{R_\text{Fr}}\int dx \exp\left(2\pi
i\left[\Phi_\text{GM} + \Phi_\text{DM}(x) \pm \Phi_B(x)\right] \right)
\end{equation}
where $f$ is the observed frequency and $x_s$ is the position of the
pulsar in the source plane.

For Figure~\ref{fig:simulation}, we selected a particular time when a
clear caustic formed, and then reran the inverse Fourier transform
twice, once adding a uniform magnetic field and once one that had a gradient.

\begin{figure}
\begin{minipage}{\linewidth}
	\includegraphics[width=\columnwidth]{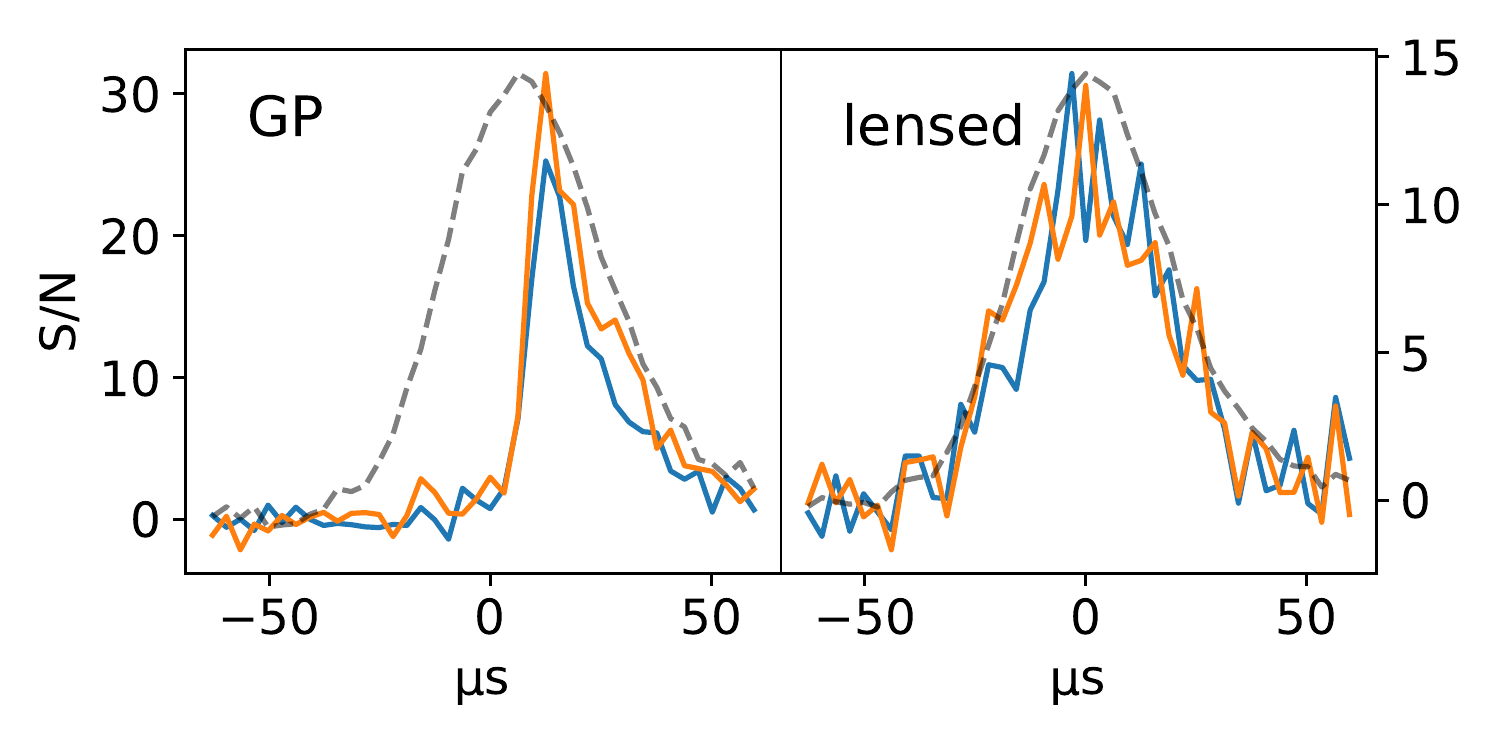}
    \vspace{-0.25cm}
\end{minipage}
\begin{minipage}{\linewidth}
	\includegraphics[width=\columnwidth]{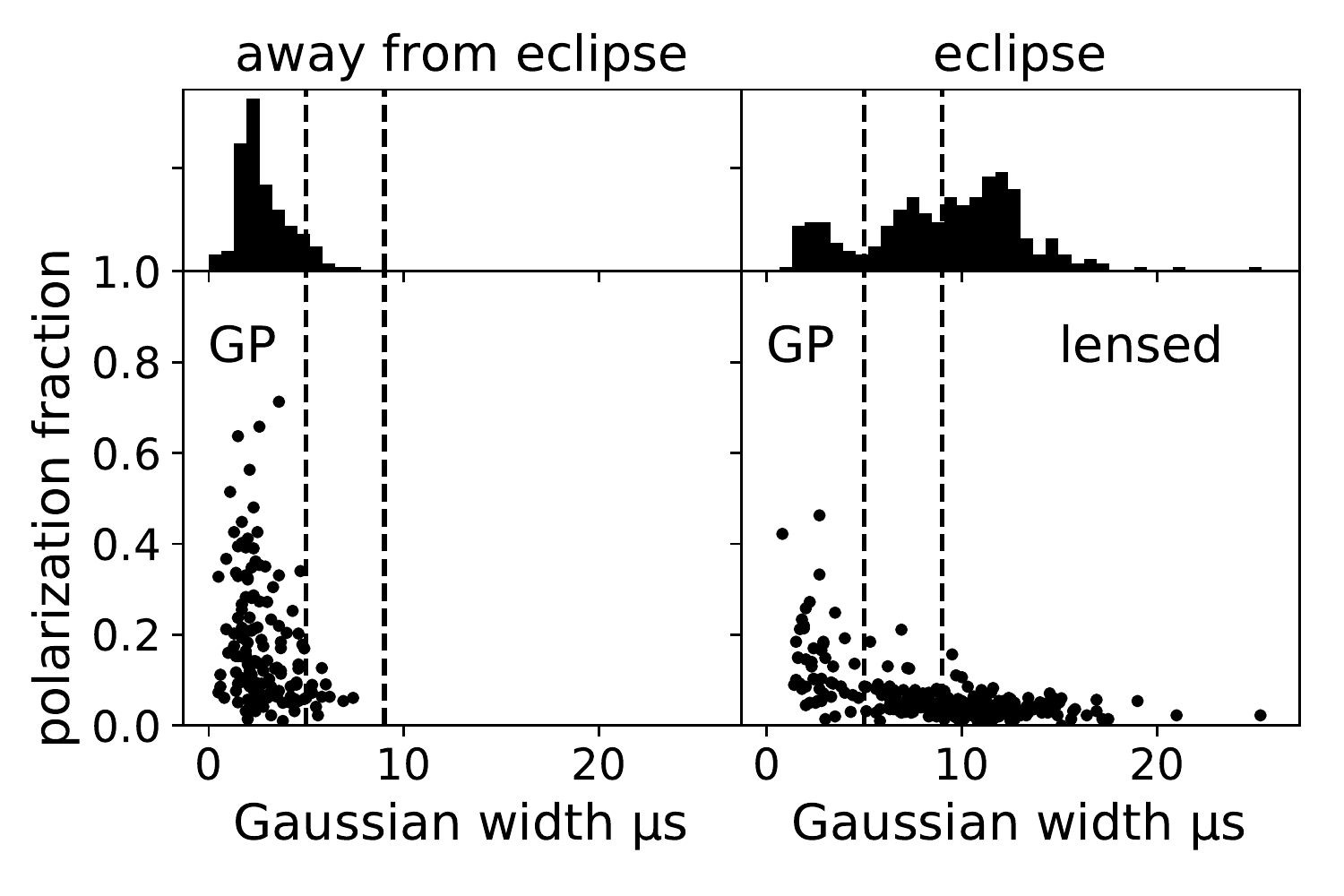}
    \vspace{-0.35cm}
\end{minipage}
\caption{Separating lensed and giant pulses.
  {\em Top:\/} The profiles of a main pulse magnified by a plasma lens
  and a giant pulse appearing at similar phase during the egress of
  the eclipse, in left and right circular polarization (colour lines),
  with the average profile integrated over the surrounding 8\,s
  overdrawn (dashed, scaled to match in amplitude).
  {\em Bottom:\/} Polarized fraction and widths of strong pulses far
  and near eclipse, with the width inferred from fitting a Gaussian
  convolved with an exponential with fixed timescale.  We consider
  pulses with fitted Gaussian width below $5\,\mu$s to be giant pulses
  and those with widths above $9\,\mu$s to be lensed ones (as
  indicated by the two vertical lines).  Note that giant pulses can be
  highly polarized, while lensed pulses are not.
}
\label{fig:sep}
\end{figure}

\section{Separating lensed and giant pulses}
\label{sec:distinguish}
In the region where the excess dispersion exceeds $10^{-4}\,\DMunit$,
we detect 268 pulses at $\mr{SN}>20$ in the main pulse phase, which
corresponds to magnification of $>10$.  The brightest pulses
clearly draw from two populations (Figure~\ref{fig:sep}, top panel):
one in which the widths are similar to that of the average main pulse
profile -- these are lensed and only appear near eclipse -- and
another in which the pulses have narrow exponential profiles -- which
appear randomly at all orbital phases.  Clearly, the first group
consist of lensed pulses, while the latter are giant pulses.

Since away from eclipse, all the bright pulses are giant pulses, we
can use the properties of those pulses as a reference in separating giant
and lensed pulses.  To obtain a quantative measure of pulse width, we
fit the profiles of all single strong pulses we detect with a Gaussian
of variable amplitude and width convolved with an exponential function
that has its timescale fixed to the scattering time of $12.2\,\mu$s
measured by \citet{17Main}.

As shown in Figure~\ref{fig:sep} (lower panel), the fitted widths for
pulses away from eclipse are all below $9\,\mu$s, and the majority of
them are below $5\,\mu$s.  Typically, they are strongly polarized
(unlike the regular pulse emission, which is nearly unpolarized;
\citealt{90Fruchter}).  For the bright pulses near eclipse, in
constrast, a much broader distribution is found, with a clear set of
short, polarized pulses and another one of wider, low-polarization
pulses.  We identify the pulses with width $<5\,\mu$s as giant pulses,
and those with width $>9\,\mu$s and low polarization fraction as
lensed pulsed.  For pulses with intermediate widths, we cannot exclude
that some are giant pulses, and hence we did not include them in our
analysis.

\bsp	
\label{lastpage}

\end{document}